%% file: Conference.tex
\documentclass[sigconf]{acmart}
\usepackage{hyperref}
\usepackage{latexsym}
\usepackage[T1]{fontenc}
\usepackage{microtype}
\usepackage{multirow}
\usepackage{graphicx}
\usepackage{adjustbox}  
\usepackage{microtype}
\usepackage{enumitem}
\usepackage{booktabs}
\usepackage{tcolorbox}          
\tcbuselibrary{breakable,skins} 
\usepackage{enumitem}
\usepackage{xcolor}             
\usepackage{amsmath}            
\usepackage{soul} 
\usepackage{cuted}
\usepackage{caption}

\newlength{\beforesecskip}
\newlength{\aftersecskip}
\newcommand{\methodname}{PushDualGen}
\AtBeginDocument{%
  }

\setcopyright{acmlicensed}
\copyrightyear{2018}
\acmYear{2018}
\acmDOI{XXXXXXX.XXXXXXX}
\acmConference[Conference acronym 'XX]{Make sure to enter the correct
  conference title from your rights confirmation email}{June 03--05,
  2018}{Woodstock, NY}
\acmISBN{978-1-4503-XXXX-X/2018/06}

\begin{document}




\title{PushDualGen: Enabling LLMs to Generate Semantic IDs with Interpretable Copy for Industrial Push Recommendation}




\author{Manjia Lin \\ Da Li \\ Yan Wang}
\affiliation{
  \institution{Kuaishou Technology, Beijing, China}
  \city{}
  \country{}
}
\email{linmanjia@kuaishou.com}



\author{Yong Jin \\ Zheming Ding \\Wei Yuan}
\affiliation{
  \institution{Kuaishou Technology, Beijing, China}
  \city{}
  \country{}
}
\email{jinyong@kuaishou.com}

\author{Lei Yan \\ Yanan Xia \\ Lu Zhang}
\affiliation{
  \institution{Kuaishou Technology, Beijing, China}
  \city{}
  \country{}
}
\email{yanlei08@kuaishou.com}

\author{Fan Yang}
\affiliation{
  \institution{Kuaishou Technology, Beijing, China}
  \city{}
  \country{}
}
\email{yangfan@kuaishou.com}

\author{Xuanping Li}
\affiliation{
  \institution{Kuaishou Technology, Beijing, China}
  \city{}
  \country{}
}
\email{lixuanping@kuaishou.com}

\author{Yanan Niu}
\affiliation{
  \institution{Kuaishou Technology, Beijing, China}
  \city{}
  \country{}
}
\email{niuyanan@kuaishou.com}

\renewcommand{\shortauthors}{Lin et al.}

\begin{abstract}
Push recommendation in \textbf{KuaiShou} proactively delivers personalized content to nearly one billion users to facilitate their engagement. 
Recently, generative recommendation has achieved end-to-end user personalization through semantic ID. However, their black-box characteristics make recommendation logics difficult to trace, hindering their deployment. OneRec-Thinking addresses this by incorporating CoT before generating SIDs, but this significantly increases inference cost. 
To support large-scale industrial applications, we propose \textbf{\methodname}, a lightweight generator, which first generates the SID and then produces a copy as a skippable explanation. \methodname~ has been deployed in Kuaishou's push recommendation system. 
Online A/B tests demonstrate the effectiveness of \methodname, delivering significant improvements in both user attraction and satisfaction. The effective play rate for videos recommended to users has relatively increased by 8.50\%, while the dissatisfaction rate has relatively fallen by 37.70\%.
In the long term, \methodname~optimises the content ecosystem, providing more exposure for long-tail videos.


\end{abstract}

\ccsdesc[300]{Information systems~Recommender systems}

\keywords{Push Notification, Generative Recommendation, Semantic ID}

\maketitle



\input{Recsys_Sections/Intro}
\input{Recsys_Sections/Related_Work}

\input{Recsys_Sections/Method}
\input{Recsys_Sections/Experiment}
\input{Recsys_Sections/Analysis}

\input{Recsys_Sections/Case_Study}
\input{Recsys_Sections/Conclusion}

\clearpage
\bibliographystyle{ACM-Reference-Format}
\bibliography{recsys_base}





\clearpage
\appendix
\input{Recsys_Sections/Appendix}

\end{document}

%% file: Recsys_Sections/Intro.tex
\section{Introduction}
\label{sec:intro}

Push recommendations utilise notifications to recommend content of interest to users without requiring app entry~\cite{nrtnotif,nvco,sistn,tim}. Push serves as a primary re-engagement channel, reaching nearly one billion users on Kuaishou. Unlike passive feed recommendations, push notifications reach users proactively. Any misaligned recommendation thus risks directly triggering user dissatisfaction, imposing stricter requirements on item selection accuracy and traceability.


Recently, the generative recommendation paradigm based on the semantic ID (SID)~\cite{tiger}, such as OneRec~\cite{onerec,onerec_report_v2,open_onerec}, has demonstrated competitive performance across a range of applications. Unlike traditional methods, they learn user preferences in an end-to-end manner and generate the corresponding semantic IDs(SIDs) for content of interest to users. OneRec-Thinking~\cite{onerecthink} incorporates a chain-of-thought before generating the SID, enhancing performance while providing the reasons for the recommendation. This provides a promising solution for the push recommendation, enabling the platform to deliver precise and reliable recommended content for user engagement. However, considering the efficiency and the throughput requirements, it is difficult to deploy in the push recommendation.

In this work, we propose \textbf{\methodname}, a generative recommendation framework tailored for push notifications. It includes three stages: (1) \emph{Semantic ID-enabled Context Compression} to encode videos into the SIDs and train LLMs to use these SIDs; (2) \emph{Personalized SID and Copy Generation} enables LLMs to generate the SID first, and then selectively produce copy as an explanation; (3) \emph{Representation Fusion for Online Service}. 
Currently, \methodname~is deployed in the push recommendation service of Kuaishou, serving nearly a billion users with approximately 100K QPS. 
Online A/B testing across 150 million users demonstrated that \methodname~delivers direct performance gains for the platform. 
Furthermore, \methodname~optimises the content ecosystem, providing exposure for long-tail videos and enhancing the attraction of the Kuaishou.

%% file: Recsys_Sections/Related_Work.tex
\section{Related Work}
\label{sec:related}

\subsection{Push Recommendation}
\label{sec:notification}
Push recommendation determines what content to deliver to users. Traditional methods retrieve candidates according to the atomic-level scoring. TIGER~\cite{tiger} proposes generative retrieval paradigm by reframing candidate generation as an autoregressive sequence generation problem over SIDs. OneRec~\cite{onerec,onerec_report_v2,open_onerec} scales this paradigm for large-scale industrial deployment using a unified generative model and further incorporates iterative preference alignment to improve generation quality. OneRec-Think~\cite{onerecthink} further augments SID generation with chain-of-thought; the added decoding overhead is prohibitive for latency-sensitive push scenarios~\cite{atspeed,l2d}. And RecGPT~\cite{recgpt,recgptv2} improves semantic coverage by augmenting online retrieval with LLM-predicted item tags as an additional retrieval signal.

\subsection{Semantic ID for Recommendation System}
\label{sec:genrec}
Generative recommendation formulates the recommendation task as a sequence generation problem, where each item is represented by a SID, a structured and discrete codeword sequence obtained from continuous item embeddings via quantization for generation.
Recent studies, such as the OneRec series~\cite{onerec, onerec_report_v2, open_onerec}, have demonstrated the effectiveness of SIDs in generative recommendation. Most existing methods construct SIDs with Residual Quantization (RQ), which hierarchically quantizes item embeddings to capture fine-grained semantics. However, RQ suffers from error accumulation: each level quantizes only the residual from the previous stage, making deeper SID tokens increasingly noisy and unreliable~\cite{parallelsid}. To alleviate this issue, several works have explored parallel SID generation~\cite{diffgrm, parallelsid, lladarec}. Nevertheless, these methods still derive SIDs from a single embedding.

%% file: Recsys_Sections/Method.tex
\begin{figure*}[htbp!]
  \centering
  \includegraphics[width=0.95\linewidth]{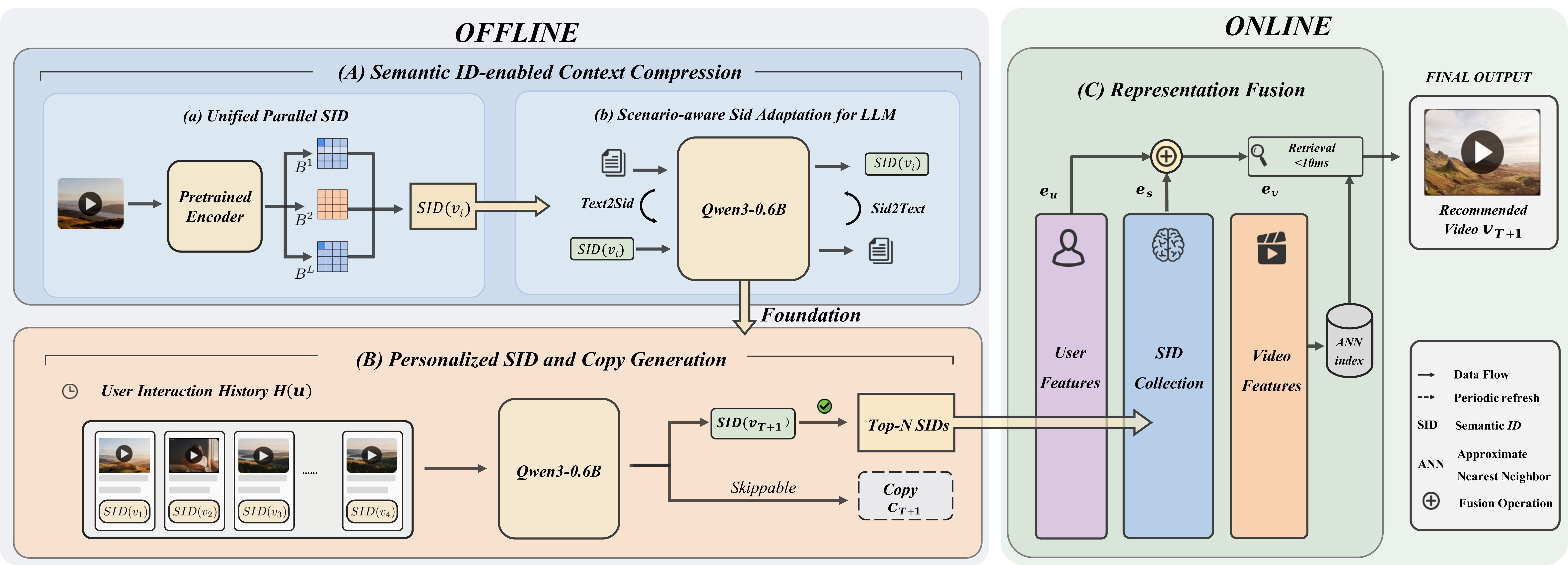}
\caption{Overview of \methodname. Module (A) compresses videos into Parallel SID and aligns them with the token of LLMs via Text2SID/SID2Text fine-tuning. Module (B) generates SIDs of videos that users may be interested in and skippable copy from the interaction history. Module (C) fuses LLM-derived preference signals with user features for ANN search.}
  \label{fig:framework}
\end{figure*}

\section{Method}
\label{sec:method}

\subsection{Semantic ID-enabled Context Compression}\label{sec:psid}
The core of push recommendation is to infer user preferences from interaction history. However, user behavioral sequences often exceed the context window of LLMs, while heuristic strategies such as truncation or random sampling may discard signals. To make full use of the input context, we replaced the original input with SIDs inspired by the parallel embedding learned from CREM~\citep{crem}.

\textbf{Parallel Semantic ID.}
Unlike residual SID methods, which suffer from hierarchical dependency and error accumulation, parallel SID approaches derive all SID tokens from a set of parallel embedding slots rather than from a single dense item representation. We construct SIDs from multiple parallel embedding slots. The details of the parallel representation model are described in Appendix~\ref{sec:parallel model}.

Specifically, each video $v$ is represented by $M$ embeddings $\{\mathbf{e}^{(m)}\}_{m=0}^{M-1}$. We quantize each embedding independently using K-means clustering. This yields $M$ codebooks $\{\mathcal{C}^{(m)}\}_{m=0}^{M-1}$, each containing $K$ centroids. For slot $m$, the SID token is obtained by nearest-centroid assignment:
\begin{equation}
s^{(m)}(v)=arg\,min_{k\in\{0,\ldots,K-1\}}
\|\mathbf{e}^{(m)}(v)-\mathbf{c}^{(m)}_k\|^2,
\end{equation}
where $\mathbf{c}^{(m)}_k \in \mathcal{C}^{(m)}$ is the $k$-th centroid of the $m$-th codebook. The final SID obtained through quantization is defined as
\begin{equation}
\mathrm{SID}(v)=\left[s^{(0)}(v), s^{(1)}(v), \ldots, s^{(M-1)}(v)\right].
\end{equation}

By replacing each video with its SID, we compress long behavioral histories into compact, discrete semantic sequences, allowing the LLM to retain more interaction history within a limited context. In practice, we set $M=8$ and $K=512$. We adopt K-means due to its simplicity and efficiency. Exploring more advanced quantization strategies is left for future work.

\textbf{Scenario-Aware SID Adaptation.}
Since SIDs are discrete codebook indices absent from the LLM vocabulary, LLMs cannot understand or generate them without alignment. We address this in two steps. First, we register all $L{\times}K$ codebook indices as special tokens with random initialization.
Second, we train LLMs via two complementary tasks over video $v_i$ with push copy $c_i$ and description $d_i$.
We begin by utilising the text $c_i$ and $d_i$ to predict the corresponding SID of video $v$ (denoted as \texttt{<T2S>}), compressing scenario-related information into the discrete SID. Simultaneously, we designed the symmetrical task, utilising the SID of the video to generate $c_i$ and $d_i$ (denoted as \texttt{<S2T>}), enabling LLMs to learn to utilise the semantic information contained in SIDs:
\begin{equation}
\begin{aligned}
&s_i^0\,s_i^1\,\cdots\,s_i^{L-1} = \mathcal{G}(c_i,\,d_i,\,\texttt{<T2S>};\,\theta), \\
&c_i\,d_i = \mathcal{G}(s_i^0,\,s_i^1,\,\cdots,\,s_i^{L-1},\,\texttt{<S2T>};\,\theta),
\end{aligned}
\nonumber
\end{equation}
where $s_i^l$ denotes the SID of $v_i$ in the $l$-th codebook, with $l \in \{0, \ldots, L-1\}$. Both tasks are jointly optimized on Qwen3-0.6B~\cite{qwen3} via next-token prediction:
\begin{equation}
\mathcal{L}_{\text{adapt}} = -\sum_{m=0}^{M-1} \log P(y^{m} \mid x,\, y^{<m}),
\end{equation}
where $x$ denotes the input and $y^m$ is the $m$-th target token.
To ensure training efficiency and stability, we freeze the original vocabulary embeddings and optimize only the SID token embeddings.

\subsection{Personalized SID and Copy Generation}
\label{sec:personalized_sft}

After compression, each video can be encoded as a parallel SID, and the generator gained the ability to understand and generate SIDs. 
Based on these foundations, we fine-tune the generator using the interaction history of user $u$, enabling it to generate the SID and the copy corresponding to the video of interest to the user.
In addition to the interaction history, the input $\mathcal{X}(u)$ also includes a task instruction, the user profile, and temporal recency markers.

\textbf{Training Objective.} 
We train the generator to generate SIDs first, and then produce the corresponding copy. To separate them, we insert a special token $\langle id2text\_sep \rangle$ into the middle. Consequently, the training objective is divided into two parts. 
The first is the prediction of the SID, formalised as follows:
\begin{equation}
  \mathcal{L}_{SID}
  = -\sum_{l=1}^{L}
    \log P\bigl(s_{T+1}^l \mid
    \mathcal{X}(u),\,
    s_{T+1}^{<l}\bigr)\,.
  \label{eq:nip}
\end{equation}
The second is the generation of copy based on the predicted SID:
\begin{equation}
\mathcal{L}_{Copy}
  = -\sum_{j=1}^{|c_{T+1}|}
    \log P\bigl(t_j \mid
    \mathcal{X}(u),\,
    \mathrm{SID}(v_{T+1}),\,\langle id2text\_sep \rangle\,, t_{<j}\bigr)\,.
  \label{eq:copy}
\end{equation}
The overall objective is the weighted sum of them:
\begin{equation}
  \mathcal{L}_{gen}
  = \mathcal{L}_{SID}
    + \lambda_{Copy}\,\mathcal{L}_{Copy}\,,
  \label{eq:stage2}
\end{equation}
where $\lambda_{Copy}$ is a hyperparameter used to balance these different training objectives.

\textbf{Multi-Token Binding.}
Although SIDs compactly encode video, the input also contains a large amount of text, which results in a high computational cost. We address this by merging frequent token sequences into single tokens, which are sourced from high-frequency $n$-grams ($n\!\in\!\{2,3,4\}$) and semantically atomic entities such as SID. And each new token embedding is initialized by mean pooling over its constituents.

\subsection{Representation Fusion for Online Serving}
\label{sec:three_tower}

We train \methodname~based on click history. Although the click contains direct signals of interest, it can only partially reflect users’ preferences; user interests are widely distributed across different situations, such as browsing and commenting. 
We incorporate the SIDs generated by \methodname~into the original user feature and fuse them with heterogeneous online features to yield a more comprehensive preference representation of the user. 
Specifically, we use three independent encoders to project (1) user features into $\mathbf{e}_u \in \mathbb{R}^d$, (2) candidate video features into $\mathbf{e}_v \in \mathbb{R}^d$, and (3) the Top-$N$ SIDs into $\mathbf{e}_s \in \mathbb{R}^d$, which summarizes the preference signal captured in \methodname. In practice, we set $N{=}20$. To incorporate the generative signal into the user representation, we fuse $\mathbf{e}_u$ and $\mathbf{e}_s$ via a weighted combination, formalised as follows:
\begin{equation}
  \mathbf{e}_u' = \alpha\,\mathbf{e}_u + \beta\,\mathbf{e}_s,
  \label{eq:fusion}
\end{equation}
where $\alpha$ and $\beta$ are hyperparameters, which are set to 1 in applications. $\mathbf{e}_u'$ is used to search for videos of interest to users from a collection consisting of $\mathbf{e}_v$, using ANN search.

%% file: Recsys_Sections/Experiment.tex
\section{Experiments}
\label{sec:experiments}
\subsection{Setup}
\label{sec:exp_setup}
\textbf{Training Data.}
Our training data is derived from the push-click logs of Kuaishou, which are collected from nearly a billion users and span millions of candidate videos. 
For \emph{SID-enabled Context Compression}, we use over 4 million videos sampled from the content pool. We filtered various types of videos to ensure content coverage and diversity. Finally, the corpus contains approximately 0.72B tokens.
For \emph{Personalized SID and Copy Generation}, we conduct incremental training based on user click logs with approximately 3.6B tokens every day.

\textbf{Implementation Details.} 
We instantiate the three components of \methodname~as follows. For \emph{Semantic ID-enabled Context Compression}, we adopt Qwen2.5-Omni-3B~\citep{qwen25_omni} with 8 learnable compression tokens to map the representation of videos into SIDs. For \emph{Scenario-aware SID Adaptation} and \emph{SIDs and Copy Generation}, we employ Qwen3-0.6B~\citep{qwen3} as the backbone generator. The two parts are trained with learning rates of \(1 \times 10^{-5}\)and \(1 \times 10^{-7}\) for 3 epochs and 1 epoch, respectively.


\textbf{Evaluation.} We conduct a 14-day A/B test by allocating 15\% of platform traffic, covering approximately 150M users, with an equal split between \methodname~and the online service based on a cascading pipeline. To mitigate potential pre-experiment imbalance introduced by hash-based traffic partitioning, we estimate treatment effects using CUPED~\citep{isoce}. We evaluate performance using four metrics. \textit{Click PV} and \textit{DAU} are used to measure notification engagement, and \textit{Eff. Play Rate} (Effective Play Rate) and \textit{Dis. Rate} (Dislike Rate) are used to evaluate post-click satisfaction in terms of content relevance and explicit negative feedback.

\begin{table}[htbp!]
\caption{Online A/B Test Results. M denotes million. Improvements significant at $p < 0.05$ are marked with $^\ast$.}
\label{tab:funnel} 
\centering
\resizebox{0.99\linewidth}{!}{
\begin{tabular}{lccr}
\toprule
&  & \textbf{Online Service} & \textbf{\methodname} \\
\midrule
\multirow{2}{*}{\textbf{\textit{Notification}}}
    & \textit{Click PV} $\uparrow$  & 82.04M    & \textbf{82.39}M\,($+0.43\%^\ast$) \\
    & \textit{DAU} $\uparrow$       & 414.89M   & \textbf{415.08}M\,($+0.05\%^\ast$) \\
\midrule
\multirow{2}{*}{\textbf{\textit{Post-Click}}}
    & \textit{Dis. Rate} $\downarrow$ & $0.053\%$  & \textbf{0.033}$\%$\,(-37.70$\%^\ast)$ \\
    & \textit{Eff. Play Rate} $\uparrow$ & $64.46\%$  & \textbf{69.94}$\%\,(+8.50\%^\ast)$ \\
\bottomrule
\end{tabular}}
\end{table}

\subsection{Online A/B Test Results}\label{sec:online_results}
We evaluate \methodname~in the online push recommendation system through a controlled A/B test. And the results are shown in Table~\ref{tab:funnel}. We can find that \methodname~achieves statistically significant improvements across all metrics. 
For notification, the gains in Click PV and DAU indicate that \methodname~ effectively attracts users back into Kuaishou and the user behaviors after clicking further confirm that the pushed content aligns with user interests. In post-click, Dis. Rate is relatively reduced by $37.70\%$, and Eff. Play Rate is relatively improved by $8.50\%$. These suggest that \methodname~ improves users' satisfaction, and reduces negative feedback. 





\subsection{Ablation Study}
\label{sec:ablation}
To analyse the contributions of components in \methodname, we use the task of SID prediction to evaluate the performance of \methodname~ and its variants. 
And Pass@k is introduced to measure the probability that at least one of $k$ independently sampled SIDs matches the ground-truth SID of a user-interested item. 
We construct these variants by replacing or disabling modules one by one and retraining the model. For the variant w/o Parallel SID, we replace the parallel SID with the SID used in OneRec\cite{onerec}. And for the variant w/o SID Adaptation, we retrained Qwen3-0.6B to generate SIDs and text corresponding to videos directly. We also analysed the training strategies of Token Freeze and Multi-token Binding. Results are shown in Table~\ref{tab:ablation}. Furthermore, we show the performance of Qwen3-0.6B after training on the SID used by OneRec~\citep{onerec}.

\begin{table}[htbp!]
\centering
\caption{Results on SID Prediction. The best and optimal results are shown by \textbf{Bold} and \underline{underlined}, respectively.}
\label{tab:ablation}
\resizebox{0.85\linewidth}{!}{
\begin{tabular}{lccc}
\toprule
 & Pass@1 & Pass@4 & Pass@20 \\
\midrule
\methodname            & \textbf{0.335} & \textbf{0.391} & \textbf{0.422} \\
\quad \textit{w/o Token Freeze}        & \underline{0.333} & \underline{0.386} & 0.412 \\
\quad \textit{w/o Multi-token Binding} & 0.332 & 0.384 & \underline{0.416} \\
\quad \textit{w/o SID Adaptation} & 0.319 & 0.352 & 0.366 \\
\quad \textit{w/o Parallel SID}        & 0.326 & 0.379 & 0.401 \\
Qwen3-0.6B             & 0.312 & 0.347 & 0.358 \\
\bottomrule
\end{tabular}}
\end{table}

We found that without the training phase of SID Adaptation, \methodname~exhibited a significant drop in performance. This is because LLMs have never encountered SID-related data during their previous training, so an additional training process is necessary to enable them to understand and use SIDs. In contrast, the token freeze and multi-token binding primarily optimize the training process, so the performance gains are relatively slight. Compared with the results of the vanilla Qwen3-0.6B, each module we have introduced has had a positive impact on \methodname, and they can be combined for further performance improvements.


%% file: Recsys_Sections/Analysis.tex
\section{Impact on Content Ecosystem}
\label{sec:analysis}

\begin{figure}[htbp!]
  \centering
  \includegraphics[width=0.6\linewidth]{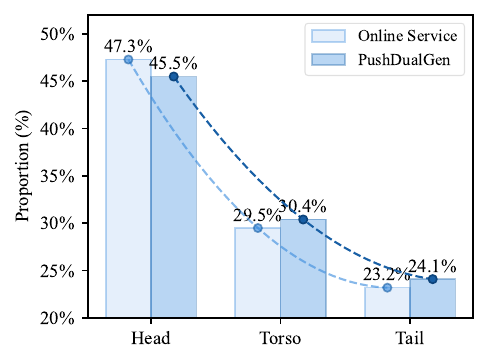}
  \caption{ The proportion of exposure among different groups. \methodname~offers fairer content exposure than the online service and is friendly to long-tail content.
  }
  \label{fig:ecosystem}
\end{figure}

For content platforms, sustained core competitiveness relies on the long-term health of the entire content ecosystem. In this part, we analyse how \methodname~ reshapes the content ecosystem.
Based on their content, videos shared on the Kuaishou platform are automatically categorised into 39 different categories. We grouped the 39 categories into three groups: \textbf{Head} (>5\%), \textbf{Torso} (1\%--5\%), and \textbf{Tail} (<1\%) according to their share of total exposure.
As shown in Figure~\ref{fig:ecosystem}, \methodname~adjusts the exposure distribution across different groups on the platform. This allows some long-tail videos to be recommended to users on Kuaishou. From a long-term perspective, this helps the platform attract more users to share videos. 
This also demonstrates that \methodname~provides additional results from different preferences for the online service. Unlike online services that tend to favour historically popular content, the SIDs generated by \methodname~offer better generalisation capabilities and perform better in cold-start scenarios.



%% file: Recsys_Sections/Case_Study.tex
\section{Case Study} 
\label{sec:case_study}
\begin{figure}[htbp!]
  \centering
  \includegraphics[width=0.8\linewidth]{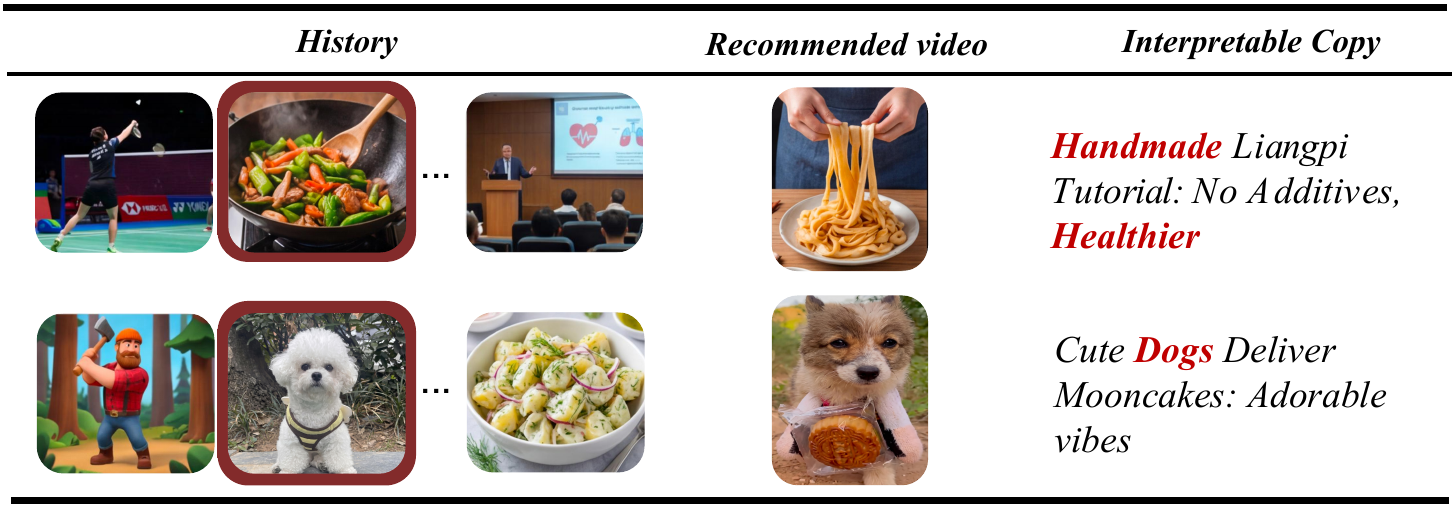}
  \caption{Qualitative analysis. Each row represents a user. \textcolor{red}{Red} indicates the link between the recommended video and the interaction history. }
  \label{fig:case-picture}
\end{figure}

\methodname~retains the ability to generate copy then SIDs. This helps us to understand what is contained within SIDs, as well as the recommendation logic behind \methodname. 
To gain a clear understanding, we conducted qualitative studies on representative examples. In addition to the recommended videos, we allow \methodname~to generate subsequent copy. We present these results in Figure~\ref{fig:case-picture}. It is clear that \methodname~captures the users' preferences effectively, which is capable of adaptively taking into account user preferences manifested through a single video click, as well as across the entire click history. 

%% file: Recsys_Sections/Conclusion.tex
\section{Conclusion}
In this paper, we introduced \methodname, a lightweight generative framework for push recommendation. 
Unlike OneRec-Thinking, \methodname~first generates SIDs and then produces subsequent copies as skippable interpretations. This design enhances performance while ensuring that the reasoning logic remains interpretable. Moreover, its flexible inference enables \methodname~to be broadly deployed.
In the future, we plan to experiment with increasing the size of the \methodname~, while expanding the application of \methodname~to scenarios beyond push recommendations.

%% file: Recsys_Sections/Appendix.tex
\section{Details of the Parallel Embedding Model and Semantic ID}\label{sec:parallel model}

\begin{figure}[htbp!]
    \centering
    \includegraphics[width=0.9\linewidth]{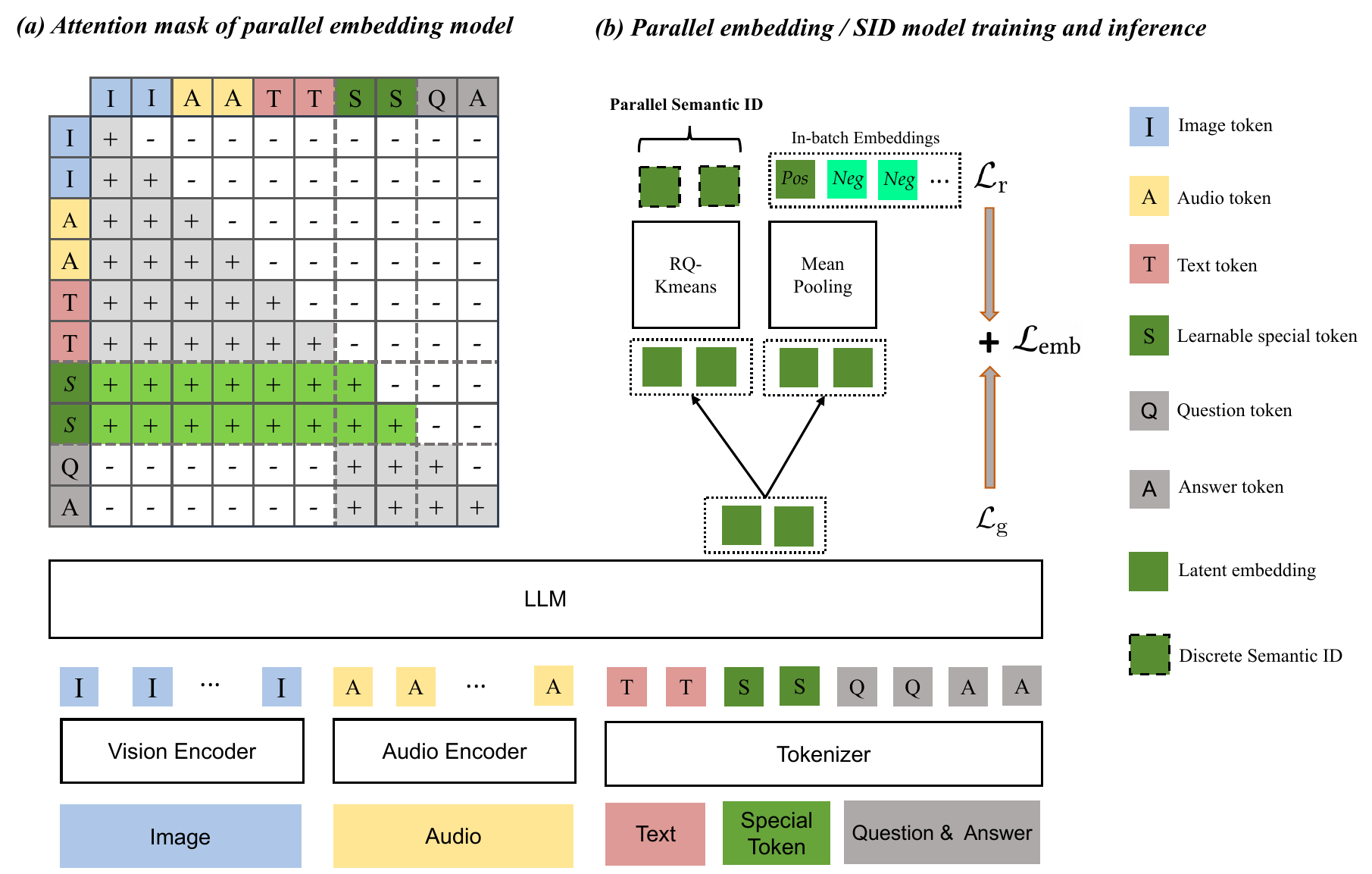}
    \caption{
        \textbf{Parallel embedding model and SID construction.}
        (a) Attention mask of the parallel embedding / SID model, where “+” indicates visible tokens and “-” indicates masked ones.
        (b) Training and inference of the parallel embedding / SID model.
    }
    \label{fig:sid_model}
\end{figure}

This section provides supplementary details of the parallel embedding model used in the main text. We build the embedding model on Qwen2.5-Omni~\citep{qwen25_omni}, following the collaborative representation learning paradigm of CREM~\citep{crem} and CoMa~\cite{coma}. The model is trained to support both video retrieval and copy generation with a shared set of embeddings. To obtain compact latent representations from multimodal inputs, we prepend a set of learnable compression tokens to the original input sequence. In our implementation, we use $8$ compression tokens, so that each video is encoded into $8$ parallel embeddings.

As illustrated in Figure~\ref{fig:sid_model}(b), the encoder is optimized with a hybrid objective:
\begin{align}
\mathcal{L}_{r}
&= - \frac{1}{B} \sum_{i=1}^{B}
\log
\frac{\exp\left(\mathrm{sim}(\mathbf{q}_i,\mathbf{p}_i)/\tau\right)}
{\sum_{j=1}^{B}\exp\left(\mathrm{sim}(\mathbf{q}_i,\mathbf{p}_j)/\tau\right)}, \\
\mathcal{L}_{g}
&= - \frac{1}{B} \sum_{i=1}^{B}\sum_{t=1}^{T_i}
\log P(x_{i,t} \mid x_{i,<t}), \\
\mathcal{L}_{\text{emb}}
&= \lambda \mathcal{L}_{r} + (1-\lambda)\mathcal{L}_{g}.
\end{align}
Here, $\mathcal{L}_{r}$ is an in-batch InfoNCE retrieval loss and $\mathcal{L}_{g}$ is the standard autoregressive language modeling loss. $B$ denotes the batch size, $\mathbf{q}_i$ and $\mathbf{p}_i$ denote a matched query--passage pair, $\mathrm{sim}(\cdot,\cdot)$ is the similarity function, $\tau$ is a temperature hyperparameter, and $\lambda \in [0,1]$ balances the two objectives. The corresponding attention mask is shown in Figure~\ref{fig:sid_model}(a).

\paragraph{Semantic ID construction.}
The resulting parallel embeddings are discretized into SIDs following the procedure described in Sec.~\ref{sec:psid}. In practice, we use $M=8$ embedding slots and $K=512$ centroids for each slot-specific codebook. We adopt K-means for its simplicity and scalability.